\documentclass[twocolumn,floatfix,superscriptaddress,longbibliography,PRL]{revtex4}
\usepackage{amsfonts,amssymb,amsmath,hyperref}
\usepackage[applemac]{inputenc}
\usepackage{multirow}
\usepackage{color}
\usepackage{dcolumn}
\usepackage{natbib}
\usepackage{bm}
\usepackage{ulem}
\usepackage{cleveref}
\usepackage{array}
\usepackage{tabularx}
\newcolumntype{Y}{>{\centering\arraybackslash}X}

\usepackage[paperwidth=210mm,paperheight=297mm,centering,hmargin=2cm,vmargin=2cm]{geometry}

\newif\ifhyper
\hypertrue
\ifhyper
\hypersetup{
citecolor = {green},
colorlinks = {true}, 
urlcolor = {blue} 
}
\fi

\newlength{\ldag}
\def\be{\begin{equation}}
\def\ee{\end{equation}}
\def\bea{\begin{eqnarray}}
\def\eea{\end{eqnarray}}
\def\bse{\begin{subequations}}
\def\ese{\end{subequations}}
\def\bc{\begin{center}}
\def\ec{\end{center}}

\newcommand{\vcrule}{\rule[-0.4cm]{0cm}{1cm}}

\newcommand{\smallvcrule}{\rule[-0.2cm]{0cm}{0.6cm}}
\allowdisplaybreaks 

\begin{document}

\title{Flat phase of quenched disordered membranes at three-loop order}

\author{S. Metayer} 
\email{smetayer@lpthe.jussieu.fr}
\affiliation{Sorbonne Universit\'e, CNRS, Laboratoire de Physique Th\'eorique et Hautes Energies, LPTHE, 75005 Paris, France}

\author{D. Mouhanna} 
\email{mouhanna@lptmc.jussieu.fr}
\affiliation{Sorbonne Universit\'e, CNRS, Laboratoire de Physique Th\'eorique de la Mati\`ere Condens\'ee, LPTMC, 75005 Paris, France}


\begin{abstract}

We study quenched disordered polymerized membranes in their flat phase by means of a three-loop perturbative analysis performed in dimension $D = 4-\epsilon$. We derive the renormalization group equations at this order and solve them up to order $\epsilon^3$. Our results confirm those obtained by Coquand {\it et al.} within a nonperturbative approach [Phys. Rev. E 97, 030102(R) (2018)] predicting a finite-temperature, finite-disorder {\it wrinkling} transition and those obtained by Coquand and Mouhanna within a recent two-loop order approach [Phys. Rev. E 103, L031001 (2021)], while correcting some of the results obtained in this last reference. We compute the anomalous dimensions that characterize the scaling  behavior at the various fixed points of the renormalization group flow diagram. They appear to be in strong agreement with those predicted within the nonperturbative context.

\end{abstract}

\maketitle 

{\it Introduction.} Understanding the effects of quenched disorder in the flat phase of polymerized membranes has become a major challenge with, as a priority target, unveiling new phenomena in the physics of graphene and graphene-like materials \cite{novoselov04,castroneto09,katsnelson12,amorim16,akinwande17}
going from mechanical ones --- e.g., a paradoxical enhancement of elasticity modulus with the density of defects \cite{lopez15} --- to electronic ones --- e.g., the possibility to open a tunable band gap \cite{liu15,yang18,ni18}. However, the interest for quenched disorder in membranes has a longer history and goes back to the early experiments of Sackman {\it et al.} \cite{sackman85} followed by those of Mutz {\it et al.} \cite{mutz91} and Chaieb {\it et al.} \cite{chaieb06, chaieb08,chaieb13,chaieb13bis} on partially polymerized lipid membranes. These authors have shown that, upon cooling below the melting temperature, these systems undergo a phase transition from a smooth structure at low-disorder, or equivalently at high
polymerization, into a wrinkled structure at high disorder, or equivalently at low polymerization.

These investigations have stimulated an important theoretical work aiming to identify precisely the nature of this weakly polymerized, wrinkled, phase that has been conjectured to coincide with a {\it glassy} phase, a phase mainly controlled by {\it disorder} fluctuations. Nelson and Radzihovsky \cite{nelson91,radzihovsky91b}, using a {\it one-loop} perturbative approach in the vicinity of the upper critical dimension $D=4$ have shown the irrelevance of a disorder acting only on the internal metric of the membrane. They have shown that the renormalization group (RG) flow was driven toward the disorder-free fixed point, called $P_4$, identified by Aronovitz and Lubensky \cite{aronovitz88} in their early RG approach of pure membranes. This result has then been confirmed by Radzihovsky and Le Doussal \cite{radzihovsky92} in the context of a leading order self-consistent screening approximation (SCSA). Morse {\it et al.} \cite{morse92a,morse92b} have then extended the one-loop study of Nelson and Radzihovsky by adding a curvature disorder to the metric one. They have confirmed the irrelevance of the disorder below $D=4$ and discovered a new, vanishing temperature, fixed point, called $P_5$. This fixed point has been identified, within the one-loop computation of Morse {\it et al.}, as being stable with respect to the disorders but unstable in the direction associated with the temperature, making the fixed point $P_5$ non-pertinent in the prospect of a glassy phase. These works have then been followed by a new approach relying on the use of the so-called nonperturbative renormalization group (NPRG) by Coquand {\it et al.} \cite{coquand18} on the metric/curvature-disordered model initially considered by Morse {\it et al.} \cite{morse92a,morse92b}. This approach is based on the use of an exact equation that controls the RG flow of an effective action $\Gamma_k$ where $k$ is a running scale. The quantity $\Gamma_k$ is truncated in powers of the field and field-derivatives while keeping nonpolynomial contributions coming from the usual perturbative parameters: coupling constants, temperature, $1/N$ --- $N$ being the number of components of the order parameter --- and so on (see \cite{kownacki09,braghin10,essafi11,hasselmann11,essafi14,coquand16a} for the use of this technique in the context of disorder-free membranes). A striking result obtained by means of this approach \cite{coquand18} is the discovery of a finite-temperature, finite-disorder, critical fixed point $P_c$, {\it unstable} with respect to the temperature, making the vanishing temperature fixed point $P_5$ {\it fully attractive} at sufficiently low temperatures. This approach has, thus, confirmed theoretically the possibility of a whole glassy phase at low temperatures in quenched disordered membranes. Moreover, the various scaling laws observed by Chaieb {\it et al.} \cite{chaieb06, chaieb08,chaieb13,chaieb13bis} in their investigations of partially polymerized lipid membranes have been qualitatively and quantitatively explained \cite{coquand20} on the basis of the analysis performed in \cite{coquand18}.

Although convincing the NPRG approach of Coquand {\it et al.} \cite{coquand18} has happened to be at odds with the results obtained from the SCSA approach of Le Doussal and Radzihovsky \cite{ledoussal18} that includes both metric and curvature disorders and in which the critical fixed point $P_c$ is missing. This is notably for this reason that, very recently, a {\it two-loop order} perturbative approach in the vicinity of $D=4$ has been performed by Coquand and Mouhanna \cite{coquand20b}, following the early one-loop order computation of Morse {\it et al.} \cite{morse92a,morse92b} and the two-loop order ones performed on disorder-free membranes by Mauri and Katsnelson \cite{mauri20} and Coquand {\it et al.} \cite{coquand20a}. This approach has confirmed the existence of a critical fixed point $P_c$ associated with a phase transition between a high-temperature phase \footnote{This phase should not be confused with the high-temperature, crumpled, phase of membranes.} controlled by the disorder-free fixed point $P_4$ and a low-temperature phase controlled by the vanishing-temperature, infinite-disorder, fixed point $P_5$. However, this approach was not conclusive as it has led to an indetermination of the coordinates of the various fixed points $P_5$ and $P_c$ as well as of the corresponding anomalous dimensions. 

In order to clarify this situation we investigate here the flat phase of quenched disorder membranes by means of a {\it three-loop} order computation in the vicinity of $D=4$, following the very recent approach performed in the pure case by Metayer {\it et al.} \cite{metayer22}; see also \cite{pikelner22} for a four-loop computation. We show that the indetermination discussed above was associated  with an incorrect expression for the RG function of the curvature disorder. We provide here the correct expressions of the RG functions at three-loop order and determine all physical quantities up to order $\epsilon^3$ without ambiguity. Our results confirm, within the perturbative context, the existence of a new fixed point $P_c$ in the flat phase of quenched polymerized membranes, even if it is found to be marginally --- to order $O(\epsilon^2)$ --- stable in contradiction with the result of the NPRG approach. However, the anomalous dimensions computed at the various fixed points $P_5$ and $P_c$ are in strong agreement with those predicted within the NPRG approach. 

{\it The action.} The action relevant to study the flat phase of quenched disordered $D$-dimensional membranes embedded in a $d$-dimensional Euclidean  space, is given by \cite{morse92a,morse92b}:
\begin{flalign}
S = \int \text{d}^Dx \ \bigg\{& \frac{{\widetilde\kappa_{\alpha\beta}}}{2} {\Delta}\bm{h}^{\alpha} ({\bf x}){\Delta}\bm{h}^{\beta} ({\bf x}) + \frac{\widetilde\lambda_{\alpha\beta}}{2} u_{ii}^{\alpha}({\bf x}) u_{jj}^{\beta}({\bf x}) \nonumber \\
& \hspace{0cm}+ {\widetilde \mu_{\alpha\beta}}\, u_{ij}^{\alpha}({\bf x}) u_{ij}^{\beta}({\bf x}) \bigg\}\ . 
\label{S}
\end{flalign}
In this expression the field $\bm{h}^{\alpha} ({\bf x})$ describes the $d_c=d-D$ flexural modes that parametrize, at each point $\bf x$ of the membrane, the height, transverse, fluctuations with respect to a fully flat configuration while $u_{ij}^{\alpha}({\bf x})$ is the strain tensor given by 
\begin{equation}
u_{ij}^{\alpha}\simeq \frac{1}{2} \left[\partial_i u_j^{\alpha}+\partial_i u_j^{\alpha}+ \partial_i {\bm h}^{\alpha} . \partial_j {\bm h}^{\alpha} \right]\ ,
\nonumber
\end{equation}
where ${\bf u}^{\alpha}$ represent $D$ longitudinal --- phonon --- modes describing the elastic, transversal, fluctuations with respect to the flat configuration. Note that in all the expressions above the Greek indices are associated with the $n$ replica that are used to performed the average over the disorder, see \cite{morse92a,morse92b,coquand18,coquand20b}. The coupling constants entering in Eq.(\ref{S}), $\widetilde \kappa^{\alpha\beta} =\widetilde\kappa \,\delta^{\alpha\beta} -\widetilde \Delta_\kappa \, J^{\alpha\beta}$, $\widetilde\mu^{\alpha\beta} = \widetilde\mu\,\delta^{\alpha\beta} - \widetilde\Delta_\mu \,J^{\alpha\beta}$ and $\widetilde\lambda^{\alpha\beta} = \widetilde\lambda \,\delta^{\alpha\beta} - \widetilde\Delta_\lambda \,J^{\alpha\beta}$, where $J^{\alpha\beta}\equiv 1$ $\forall\, \alpha,\beta$, encode the bending rigidity $\widetilde\kappa$, the elastic coupling constants --- Lam\'e coefficients --- $\widetilde\lambda$ and $\widetilde \mu$ and, finally, the short-range disorder variances $\widetilde\Delta_{\kappa}$, associated with the curvature disorder and ($\widetilde\Delta_{\lambda}$, $\widetilde\Delta_{\mu}$) associated with  the metric disorder. As usual, see \cite{morse92a,morse92b,coquand18,coquand20b}, the temperature $T$ has been re-absorbed in the definition of the coupling constants: $\widetilde{g}=\left\{\kappa/T,\lambda/T,\mu/T,\Delta_{\lambda}/T^2,\Delta_{\mu}/T^2,\Delta_{\kappa}/T^2\right\}$. Note that stability considerations imply that $\mu$, $ \lambda+2\mu/D$, $\Delta_{\kappa}$, $\Delta_{\mu}$ and $\Delta_\lambda+2\Delta_{\mu}/D$ should be all positive.

Finally we define, as in \cite{morse92a,morse92b,coquand18,coquand20b}, the relevant correlation functions. Writing $\delta{\bm h}({\bm q})={\bm h}({\bm q})-\langle{\bm h}({\bm q})\rangle$ one has, denoting by $[\dots]$the average over a Gaussian disorder and by $\langle\dots\rangle$ the thermal average: 
\begin{equation}
\begin{array}{ll}
G_{h_i h_j}({\bm q})&= \big[\langle h_i({\bm q}) h_j(-{\bm q})\rangle \big]= T \chi_{h_i h_j}({\bm q})+ C_{h_i h_j}({\bm q})
\nonumber
\end{array}
\end{equation}
where $T \chi_{h_i h_j}({\bm q})= \big[\langle \delta h_i({\bm q}) \delta h_j(-{\bm q})\rangle \big]$ and $C_{h_i h_j}({\bm q})=\big[\langle h_i({\bm q})\rangle\langle h_j(-{\bm q})\rangle\big]$ that behave, at low momenta, as: 
\begin{equation}
\chi_{h_i h_j}({\bm q})\sim q^{-(4-\eta)}, \hspace{0.5cm} C_{h_i h_j}({\bm q})\sim q^{-(4-\eta')}\ ,
\nonumber
\end{equation}
and are controlled by the exponents $\eta$ and $\eta'$ associated with  thermal and disorder fluctuations, respectively. One defines analogous correlation functions for the phonon fields $ {\bf u}$ with critical exponents $\eta_u$ and $\eta'_u$ that are related to the previous ones by Ward identities \cite{morse92a,morse92b,coquand18,coquand20b}: $\eta_u+2 \eta=4-D$ and $\eta'_u+2\eta'=4-D$. Finally, a relevant quantity obtained from $\eta$ and $\eta'$ is the exponent $\phi$ given by \cite{morse92a,morse92b} $\phi=\eta'-\eta$ that determines which kind of fluctuations dominates at a given fixed point: $\phi>0$, respectively $\phi<0$, corresponds to a fixed point where thermal, respectively disorder, fluctuations dominate. For $\phi=0$ both kinds of fluctuations coexist; the corresponding fixed point is then said to be marginal. 

\vspace{0.2cm}
{\it The renormalization group equations and fixed points.} We have derived the three-loop order RG equations for model (\ref{S}) within the modified minimal subtraction scheme and in the massless case, following the procedure used in \cite{metayer22}. In particular, we have made intensive use of techniques of massless Feynman diagram calculations, see, e.g., the review \cite{Kotikov:2018wxe}, and have recourse to QGRAF \cite{Nogueira:1991ex} for the generation of the diagrams, {\it Mathematica} to perform the numerator algebra and LiteRed \cite{Lee:2012cn,Lee:2013mka} to reduce the loop integrals to a finite set of master integrals. As in the disorder-free case \cite{metayer22} we had to evaluate 51 diagrams; here their evaluation is complicated by the heavy algebra associated with  the replica structure. Computations, that are based on an extension of the formalism developed in the pure case (see \cite{metayer22} and \cite{metayer22c}) will be detailed in a forthcoming publication, see \cite{metayer22b}. In this paper, we focus on the analysis of the RG functions. The latter are too long to be displayed in the main text; they are given in Appendix \ref{beta} up to two-loop order while the full three-loop expressions are given in a supplementary material  \footnote{The supplementary material file SM.m, containing the full 3-loop RG expressions, is available along with the
LATEX files. \label{supple}}. They are expressed in terms of dimensionless renormalized quantities: 
$\overline g= k^{D-4}Z^2\,Z_{g}^{-1}\,\widetilde\kappa^{-2}\widetilde g$ for $g\in\{\mu,\lambda,\Delta_{\mu},\Delta_{\lambda}\}$ 
and $\overline\Delta_{\kappa}=Z Z_{\Delta_{\kappa}}^{-1}\, \widetilde\kappa^{-1} \widetilde\Delta_{\kappa}$ where $Z_{\alpha}$ is the coupling constant renormalization for any coupling $\alpha$. $Z$ is the field renormalization defined through ${\bf h}=Z^{1/2}\widetilde\kappa^{-1/2}{\bf h}_R$ and ${\bf u}=Z\widetilde\kappa^{-1}{\bf u}_R$ where ${\bf u}_R$ and ${\bf h}_R$ are the renormalized fields. Note that with these definitions the scaling of the renormalized coupling constants with the temperature is given by $\overline\lambda\sim\overline\mu\sim T$ (that can be considered as measures of the temperature $T$), $\overline\Delta_{\kappa}\sim 1/T$, while the other coupling constants are temperature-independent. The running anomalous dimension $\eta_t$ and the running exponent $\phi_t$ are given by $\eta_t=-\partial_t \ln Z$ and $\phi_t=\eta_t'-\eta_t= \partial_t \ln\overline\Delta_{\kappa}$ with $t=\ln k$, $k$ being the renormalization momentum scale. 

\vspace{0.2cm}

We discuss the fixed points of the RG equations and the corresponding field anomalous dimensions, whose expressions are explicitly provided at three-loop order in Appendix \ref{FP}. We first recall the one-loop order result \cite{morse92a,morse92b}, then we provide our two-loop order results, while correcting those of Ref.\cite{coquand20b} and, finally, our three-loop order ones.

{\it One-loop order.} At one-loop order one has, in $D<4$, besides the Gaussian fixed point, two non-trivial fixed point \footnote{One forgets unstable fixed points with vanishing disorder.} located on the hypersurfaces $\overline\lambda/\overline\mu=\overline\Delta_{\lambda}/\overline\Delta_{\mu}=-1/3$: 

1) the disorder-free fixed point, $P_4$, for which $\overline\mu=96\pi^2\,\epsilon/(24+d_c)$, $\overline\Delta_{\mu}=\overline\Delta_{\kappa}=0$ and $\eta=\eta'/2=\phi=12\,\epsilon/(24+d_c)$. This attractive fixed point controls the long distance behaviour of both disorder-free and disordered membranes --- see below. 

2) The vanishing temperature, infinite disorder, fixed point, $P_5$, for which $\overline\mu=0$ and $\overline\Delta_{\kappa}=\infty$. Due to the infinite value of the curvature disorder at this fixed point one has to consider a special set of coupling constants $\{\overline\mu,\,\overline g_{\mu},\,\overline g_{\lambda},\,\overline\Delta_{\mu},\,\overline\Delta_{\lambda}\}$ involving $\overline g_{\mu}=\overline\mu\, \overline\Delta_{\kappa}$ and $\overline g_{\lambda}=\overline\lambda\, \overline\Delta_{\kappa}$ that stay finite at $P_5$. At this fixed point one has: $\overline\Delta_\mu= 24\pi^2\,\epsilon/d_{c6}$, $\overline g_\mu=48\pi^2\,\epsilon/d_{c6}$ and $\eta=\eta'=3\,\epsilon/d_{c6}$ with $d_{c6}=d_c+6$. At this order $\phi=0$; $P_5$ is thus marginal. An analysis of the nonlinearities of the RG flow shows that $P_5$ is marginally unstable \cite{morse92a,morse92b}.

\vspace{0.2cm}
{\it Two-loop order.} At two-loop order \cite{coquand20b} one recovers the disorder-free fixed point $P_4$ whose coordinates and anomalous dimension are given in \cite{coquand20a}. We focus here on the non-trivial fixed points with non-vanishing disorder. To identify both the vanishing temperature fixed point $P_5$ and the putative critical fixed point $P_c$ obtained by Coquand {\it et al.} \cite{coquand18} one has again to recourse to the set of coupling constants $\{\overline\mu,\,\overline g_{\mu},\,\overline g_{\lambda},\,\overline\Delta_{\mu},\,\overline\Delta_{\lambda}\}$ that we expand in powers of $\epsilon$ up to order 2, $\overline{g} = \sum_{i=1}^2 \mathcal{C}_g^{(i)}\,\epsilon^i$, where the $\mathcal{C}_g^{(1)}$'s are the coordinates of $P_5$ at one-loop order, given above. One finds, contrary to \cite{coquand20b}, 
that the coordinates of $P_5$ at two-loop order are completely determined: $\overline\mu=0$, $\overline g_\mu=48\pi^2\,\epsilon/d_{c6}+O(\epsilon^2)$, $\overline g_\lambda=-16\pi^2\,\epsilon/d_{c6}+O(\epsilon^2)$, $\overline\Delta_\mu=24\pi^2\,\epsilon/d_{c6}+O(\epsilon^2)$ and $\overline\Delta_\lambda=-8\pi^2\,\epsilon/d_{c6}+O(\epsilon^2)$. Their full expressions up to order $\epsilon^3$ as well as the corresponding anomalous dimensions $\eta_5$ and $\eta'_5$ are given in Table \ref{tablexpo1} of Appendix \ref{FP}. 

Note that, at this order, $P_5$ no longer belongs to the hypersurfaces $\overline\lambda/\overline\mu=\overline g_\lambda/\overline g_\mu=\overline\Delta_{\lambda}/\overline\Delta_{\mu}=-1/3$ by a distance of $O(\epsilon)$. This seems to be a general feature of the perturbative approach of flat phase of both disorder-free and disordered membranes beyond one-loop order. Let us give here the numerical values of the anomalous dimensions in the physical $d_c=1$~case: 
\begin{flalign}
\hspace{1.25cm}\eta_5 & =0.42857\,\epsilon-0.03779\, \epsilon^2 +O(\epsilon^3)\ , \phantom{0000000000} \nonumber \\[5pt]
\eta'_5 & =0.42857\, \epsilon-0.03341\, \epsilon^2 +O(\epsilon^3)\ , \phantom{00000000000}
\raisetag{0.75cm}
\label{eta52lnum}
\end{flalign}
and $\phi_5=0.00437\, \epsilon^2$. An important fact is that one finds $P_5$ to be {\it unstable}, in agreement with the positive value of $\phi_5$ but in disagreement with the NPRG approach. We shall come back on this fact below. 

One of the main results of our computation at two-loop order is the identification of a finite temperature, finite disorder, ``$P_c$-like", fixed point whose coordinates $\overline\mu=-8\pi^2\epsilon^2/d_{c6}^2+O(\epsilon^3)$, $\overline g_\mu=48\pi^2\,\epsilon/d_{c6}+O(\epsilon^2)$, $\overline g_\lambda=-16\pi^2\,\epsilon/d_{c6}+O(\epsilon^2)$, $\overline\Delta_\mu=24\pi^2\,\epsilon/d_{c6}+O(\epsilon^2)$ and $\overline\Delta_\lambda=-8\pi^2\,\epsilon/d_{c6}+O(\epsilon^2)$ differ, at leading order, with those of $P_5$, only by the (non-vanishing) value of $\overline\mu$. As found within the NPRG approach of Coquand {\it et al.} \cite{coquand20a} this fixed point emerges from $P_5$ as the dimension $D$ is lowered from the upper critical dimension $D=4$. The full expressions of the coordinates as well as the corresponding anomalous dimensions are given in the Table \ref{tablexpo2} of Appendix \ref{FP}. Very surprisingly $\overline\mu$ is found to be {\it negative} at leading order in $\epsilon$, which violates the condition of positivity of $\overline\mu$ required by stability considerations. Moreover, and in agreement with the unstable character of $P_5$, the fixed point $P_c$ is found to be stable, a result also at odds with those obtained within the NPRG approach. All these facts should be nevertheless considered with great care. Indeed, one knows since the two-loop perturbative approach to disorder-free membranes by Coquand {\it et al.} \cite{coquand20a} that some physical fixed points, well defined at one-loop order, are ejected from the (mean-field) region of stability or from the hypersurface $\overline\lambda/\overline\mu=-1/3$, at two-loop order. 
However the physical quantities --- as opposed to fixed points coordinates --- are trustable and in agreement with those obtained from nonperturbative (SCSA and NPRG) approaches. One gets, for the anomalous dimensions at $P_c$ for $d_c=1$:
\begin{equation}
 \eta_c= {\eta'_c}=0.42857\,\epsilon-0.03695\, \epsilon^2 +O(\epsilon^3)\ ,
\label{etac2lnum}
\end{equation}
and $\phi_c=0$ as $\partial_t\ln\overline\Delta_{\kappa}=\eta'-\eta$ vanishes at this fixed point.

One immediately sees on the series (\ref{eta52lnum}) and (\ref{etac2lnum}) a strong decrease of their numerical coefficients with the order of the expansion. The same observation was done in the context of disorder-free membranes at two-\cite{coquand20a} and three-loop \cite{metayer22} orders. This is both an indicator of the (apparent) convergence of the series and of a fruitful comparison with the nonperturbative approaches. This is done below with the three-loop order results. 

\vspace{0.2cm}
{\it Three-loop order.} At three-loop order the picture obtained at two-loop order is not fundamentally changed. One gets the disorder-free fixed point $P_4$ and two non-trivial fixed points: the vanishing temperature, infinite disorder fixed point $P_5$ and the finite disorder, finite temperature, fixed point $P_c$ whose coordinates and anomalous dimensions for any $d_c$ are given in Table \ref{tablexpo1} and \ref{tablexpo2} of Appendix \ref{FP}. Again it is more appealing to consider the anomalous dimensions in the physical $d_c=1$ case. One has: 
\begin{flalign}
\eta_5 & =0.42857\,\epsilon-0.03779\, \epsilon^2 - 0.01205\, \epsilon^3 +O(\epsilon^4)\ ,\nonumber & \\[5pt] 
\eta'_5 & = 0.42857\, \epsilon-0.03341\,\epsilon^2 - 0.00964\, \epsilon^3 +O(\epsilon^4)\ . &
\raisetag{0.75cm}
\label{eta53lnum}
\end{flalign}
One immediately notices that the coefficients associated with  the third order are still strongly decreasing with respect to those of second order. This was also observed in the disorder-free case \cite{metayer22}. The same phenomenon occurs at $P_c$:
\begin{equation}
 \eta_c= {\eta'_c}=0.42857\,\epsilon-0.03695\, \epsilon^2 - 0.01191\, \epsilon^3 +O(\epsilon^4)\ . 
\label{etac3lnum}
\end{equation}

\vspace{0.2cm}
{\it Comparison with NPRG approach.} It is very instructive to compare our results with those obtained from the NPRG approach and reexpanded in powers of $\epsilon$. The results for $\eta_5$, $\eta'_5$ and $\eta_c$ for any $d_c$ are given in Table \ref{tablexpo3} of Appendix \ref{NPRG} and, again, we consider them for $d_c=1$. For convenience all --- perturbative as well as nonperturbative --- results have been gathered in Table \ref{P1} and \ref{P2}. One first notices the structural identity between the series coming from the two approaches. Secondly, one sees that the numerical agreement is very good both at two- and three-loop orders, almost up to three digits at the fixed point $P_c$. This fact advocates for --- but does not ensure --- the identification of the fixed point $P_c$ found in this work with the one obtained within the nonperturbative context. 

\begin{table}[t]
\begin{center}
\begin{tabularx}{\columnwidth}{c|Y}
\hline
\hline
\smallvcrule
Approach & $P_5$ \\ 
\hline
\smallvcrule
&  $\eta_5= 0.42857\,\epsilon -0.03779\,\epsilon^2 - 0.01205\, \epsilon^3$ \vspace{-0.4cm} \\ 
Three-loop \vspace{-0.4cm} & \\
& $\eta'_5= 0.42857\, \epsilon-0.03341\,\epsilon^2 - 0.00964\, \epsilon^3$ 
\smallvcrule \\
\hline
\smallvcrule
& $ \eta_5= 0.42857\,\epsilon -0.03532\,\epsilon^2 - 0.01293\, \epsilon^3$ \vspace{-0.4cm} \\ 
NPRG \vspace{-0.4cm} &\\ 
& $ \eta'_5= 0.42857\,\epsilon -0.03999\, \epsilon^2 - 0.01636\, \epsilon^3$ \smallvcrule 
\\
\hline\hline 
\end{tabularx}
\end{center}
\vspace{-0.5cm}
\caption{Anomalous dimensions $\eta_5$ and $\eta'_5$ at order $\epsilon^3$ at the fixed point $P_5$ from the three-loop order approach (this work) and from the NPRG approach \cite{coquand18}.}
\label{P1}
\end{table}

\begin{table}[t]
\begin{center}
\begin{tabularx}{\columnwidth}{c|Y}
\hline \hline
\smallvcrule Approach & $P_c$ \\
\hline
\smallvcrule Three-loop & $ \eta_c= 0.42857\, \epsilon -0.03695 \, \epsilon^2 - 0.01191\, \epsilon^3$ \\
\smallvcrule NPRG  &  $ \eta_c= 0.42857\, \epsilon -0.03621 \, \epsilon^2 - 0.01318\, \epsilon^3$\\
\hline\hline 
\end{tabularx}
\end{center}
\vspace{-0.5cm}
\caption{Anomalous dimension $\eta_c=\eta'_c$ at order $\epsilon^3$ at the fixed point $P_c$ from the three-loop order approach (this work) and from the NPRG approach \cite{coquand18}.}
\label{P2}
\end{table}

\begin{table}[t!]
\begin{center}
\begin{tabularx}{\columnwidth}{Y|Y|Y}
\hline
\hline
\smallvcrule Approach & $P_5$ & $P_c$ \\ \hline 
\smallvcrule
Three-loop $\epsilon$ & $ \eta_5= 0.85714$ & $\eta_c= 0.85714$ \\ 
\smallvcrule
Three-loop $\epsilon^2$ & $ \eta_5= 0.70600$ & $\eta_c = 0.70933$ \\ 
\smallvcrule
Three-loop $\epsilon^3$ & $ \eta_5= 0.60962$ & $\eta_c = 0.61402$ \\ 
\smallvcrule
NPRG & $  \eta_5= 0.449\phantom{00}$ & $  \eta_c= 0.492\phantom{00}$ \\ 
\hline\hline
\end{tabularx}
\end{center}
\vspace{-0.5cm}
\caption{Approximations in $D=2$ of the critical exponents $\eta_5$ and $ \eta_c$ at order $\epsilon, \epsilon^2$ and $\epsilon^3$ from the perturbative approach (this work) and from the NPRG approach \cite{coquand18}.}
\label{P3}
\end{table}

Finally, on the basis of the fast decreasing character of the series giving the various anomalous dimensions one can compute, from the expressions given in Table \ref{P1} and \ref{P2}, successive approximations of $\eta_5$ and $\eta_c$ in $D=2$, {\it i.e.} taking $\epsilon=2$, without having recourse to resummation techniques that are worthless in the case of fast decreasing series. The results, displayed in Table \ref{P3}, show that the exponents $\eta_5$ and $\eta_c$ obtained in the perturbative context get closer and closer to the values obtained by means of the NPRG approach. This is only indicative but provide a further sign that the perturbative and nonperturbative techniques may describe the same physical situation.

\vspace{0.2cm}
{\sl Conclusion.} We have analyzed quenched disordered membranes by means of a three-loop order perturbative approach. We have derived the RG functions for the various coupling constants and determined the fixed points relevant to the long-distance physics of membranes at two and three-loop orders. Our findings is that it clearly exists a new finite temperature, finite disorder, fixed point $P_c$ in the RG flow diagram. The proximity between the values obtained for the anomalous dimensions at various fixed points within the perturbative and nonpertubative approaches suggests that the fixed point $P_c$ identified in the former approach coincides with that discovered within the latter one. 
Nevertheless, some difficulties encountered in the present perturbative approach prevents a straightforward conclusion into this direction. Indeed, the negative value of the coordinate $\mu_c$ and thus the corresponding negative shear modulus of $P_c$, makes this fixed point apparently both unphysical and stable while the vanishing temperature fixed point $P_5$ appears to be unstable, at odds with the results obtained within the nonperturbative context \cite{coquand18}. This is however believed to be an artefact of the present pertubative approach. Indeed, as argued in the main text, it has been shown in recent studies \cite{coquand20a,metayer22}, by some of the present authors, that in the similar case of flat phase of disorder-free membranes, analyzed perturbatively beyond one-loop order, the corrections to the fixed points coordinates can artificially eject some of them from their mean-field region of stability when extrapolated to $\epsilon=1$. As a matter of fact, this reflects the renormalization scheme and field-parametrization dependence of the RG functions beyond one-loop order for models with more than one coupling constant, see, e.g.\cite{McKeon18}, that leads to nonuniversal expressions for both RG functions and fixed points coordinates. In these models, only critical exponents, taken at fixed points can be considered as universal quantities in the pertubative context. 
We believe that the situation encountered in the present work is based on this phenomenon, reinforced by the fact that the incriminated series begin at order $\epsilon^2$, which is indicative of their misbehaved structure. Admitting these arguments, our work would be a first perturbative confirmation of the existence of a finite disorder, finite temperature transition occurring in the phase diagram of quenched disordered membranes, as well as of the existence of a low-temperature glassy phase in these systems.
However, it remains to confirm the present results by circumventing the difficulty encountered here. 
Within the perturbative context this may be achieved by changing the regularization and/or the renormalization scheme. Another method could be to enlarge the perturbative series by several orders and to use resummation techniques in order to extract reliable informations about the position of the fixed points. Unfortunately, both solutions are far beyond the scope of this paper given the complexity of the model. A more realistic approach to circumvent this issue can be found in the nonperturbative context. In this respect it would be of tremendous interest to revisit the SCSA approach in view of the present findings. 

\onecolumngrid

\acknowledgements

D. M. wishes to greatly thank O. Coquand  for discussions and M.T. for her unfailing support. S. Metayer wishes to thank his PhD supervisor, S. Teber, for his guidance and initiating him to computational techniques that have been efficiently extended to the present problem.

\appendix

\vspace*{0.25cm}

\section{Renormalization group equations at two-loop order.} 
\label{beta}

One gives here the RG equations at two-loop order for the --- dimensionless --- coupling constants (forgetting their overlining) entering in action Eq.(\ref{S}). The three-loop order contributions are too long to be displayed on paper; they are given in the supplementary material \ref{supple}. For the sake of brevity, we use the notations $\eta_{ut} = \epsilon-2\eta_t$, $a=\lambda+2\mu$, $\Delta_a=\Delta_\lambda+2\Delta_\mu$, $b_n=1+n\Delta_\kappa$, $\Theta_1=a \mu (a-\mu )$ and $\Theta_2=a^2 \Delta_\mu +\mu ^2 (a+\Delta_a)-a \mu (a+2 \Delta_\mu)$, leading:

\begin{flalign*}
& \partial_t{\mu}= -\mu\,\eta_{ut} 
+ \frac{d_c\mu^2b_2}{6(4\pi)^2}
+ \frac{d_c\mu^2}{216a^2(4\pi)^4}
\bigg[686 \Delta_\kappa ^2 \Theta_1-227 b_4 \Theta_2
\bigg]\,,
\\
&\partial_t{\lambda} = -{\lambda}\,\eta_{ut} 
+\frac{d_cb_2}{6(4\pi)^2}
\bigg[
6 a^2-18 a \mu +13 \mu^2
\bigg]-\frac{d_c}{216a^2(4\pi)^4}
\bigg[
6 d_c b_2^2 a^2 \mu (3 a-5 \mu )^2\\
&\hspace{0.9cm}-b_4 \Theta_2 (378 a^2-1674 a \mu +1819 \mu ^2)+2 \Delta_\kappa ^2 \Theta_1 (972 a^2-3726 a \mu +3641 \mu ^2)
\bigg]\,,
\\
&\partial_t{\Delta}_\mu = -{\Delta}_\mu\,\eta_{ut}
+\frac{d_c\mu}{6(4\pi)^2}
\bigg[
2 b_2 \Delta_\mu - \Delta_\kappa ^2 \mu
\bigg]
+\frac{d_c\mu}{108a^2(4\pi)^4}
\bigg[
2 \Delta_\kappa ^2 \Theta_1 (58 \mu b_{-2} +343 \Delta_\mu )+\Theta_2 (343 \mu \Delta_\kappa ^2 -227 b_4 \Delta_\mu )
\bigg]\,,
\\
& \partial_t{\Delta}_\lambda = -{\Delta}_\lambda\,\eta_{ut}
-\frac{d_c}{6(4\pi)^2}
\bigg[
\Delta_\kappa ^2 (6 a^2-18 a \mu +13 \mu ^2)-2 b_2 \big(3a(2 \Delta_a-3 \Delta_\mu) -\mu(9 \Delta_a -13 \Delta_\mu) \big)
\bigg]
\\
&\hspace{1.2cm}+\frac{d_c}{108a^2(4\pi)^4}
\bigg[
\Delta_\kappa ^2 \Big(6 d_c a^2 b_2 \mu (3 a-5 \mu )^2+a^2 \mu ^2 (4698 \Delta_a-22101 \Delta_\mu +3493 \mu )-108 a^3 \mu (18 \Delta_a-87 \Delta_\mu +19 \mu )\\
&\hspace{1.2cm} -54 a^4 (18 \Delta_\mu -7 \mu )+a \mu ^3 (14564 \Delta_\mu-1819 \mu )-3641 \mu ^4 \Delta_a\Big)-9 d_c a^2 b_2^2 (3 a-5 \mu ) (a \Delta_\mu +2 \Delta_a \mu -5 \Delta_\mu \mu )\\
&\hspace{1.2cm} +4 \Delta_\kappa ^3 \Theta_1 (297 a^2-1026 a \mu +911 \mu ^2)+b_4 \Theta_2 (378 a \Delta_a-837 a \Delta_\mu -837 \Delta_a \mu +1819 \mu \Delta_\mu )
\bigg]\,,
\\
&\partial_t{\Delta}_\kappa = 2{\Delta}_\kappa\,\eta_t
-\frac{5 \Theta_1 b_1 \Delta_\kappa}{a^2(4\pi)^2}
+\frac{ \Delta_\kappa}{72a^4(4\pi)^2}
\bigg[
a^2 \mu ^2 (b_3+2 \Delta_\kappa ^2) \Big(a^2 (39 d_c+340)-1220 a \mu +5 (212-15 d_c) \mu ^2\Big)\\
&\hspace{1.2cm}-20 \Big(a^2 \mu ^2 \Delta_\mu (45 a \Delta_\kappa +122 \Delta_a+424 \Delta_\mu )-a^3 \mu \Delta_\mu(15 a \Delta_\kappa +244 \Delta_\mu )+34 a^4 \Delta_\mu ^2\\
&\hspace{1.2cm}-a \mu ^3 \big(15 a \Delta_\kappa (\Delta_a+2 \Delta_\mu )+424 \Delta_a \Delta_\mu \big) + \mu ^4 \Delta_a(15 a \Delta_\kappa +106 \Delta_a)\Big)
\bigg]\,,
\\
&\eta_t = \frac{5}{a^2(4\pi)^2}
\bigg[
\Delta_\kappa \Theta_1-\Theta_2
\bigg]
+\frac{1}{72a^4(4\pi)^2}
\bigg[
a^2 \Big(d_c \big(\mu ^2 (39 a^2 (\Delta_\kappa b_{-3} +1)+75 \mu ^2 (3 \Delta_\kappa ^2-b_1))\big)\\
&\hspace{0.7cm}+ \mu ^2 \Delta_\kappa \big(20 (17 a^2-61 a \mu +53 \mu ^2)-90 \Delta_\kappa (13 a^2-44 a \mu +37 \mu ^2)\big)+6 (\Delta_\mu -\mu ) (d_c a b_2 \mu (13 a-25 \mu )\\
&\hspace{0.7cm}+30 a \big(10 \mu ^2 \Delta_\kappa -a (\Delta_\mu -\mu b_{-6} )\big)+150 \Theta_2)\Big)+10 a \mu \Theta_2 (244 a \Delta_\kappa +15 d_c b_2 \mu -424 \mu \Delta_\kappa )-1060 \Theta_2^2
\bigg]\,.
\end{flalign*}

\newpage

\section{Fixed points coordinates at order \texorpdfstring{$\epsilon^3$}{ep3} from three-loop approach.}
\label{FP}

We provide here the fixed points coordinates obtained from the three-loop approach (this work) for both $P_c$ and $P_5$, at order $\epsilon^3$.

\vspace{0cm}

\begin{table}[h!]
\begin{center}
\begin{tabular}{|c|l|}
\hline \smallvcrule
$\mu_5$ & \hspace{0.25cm} $ 0 $ \\ 
\hline \vcrule
${g_\mu}_5$ & $\cfrac{48\pi^2 }{\, d_{c6}}\, \epsilon-\cfrac{4\pi^2(52\, d_c^2+573\, d_c+486)}{5\, d_{c6}^4}\, \epsilon^2 -
\cfrac{\pi^2}{500 \, d_{c6}^7} \Big(31780 \, d_c^5-d_c^4 \big(4130687 - 3558816 \,\zeta_3\big) $ \vspace{-0.1cm} \\ 
& $ \hspace{6.2cm}-6\, d_c^3 \big(7627163 - 6225984 \,\zeta_3\big) - 12\, d_c^2 \big(1161793 - 1236384 \,\zeta_3\big) $ \\ 
& $ \hspace{6.2cm} + 72\, d_c \big(13807837 - 10855296 \,\zeta_3\big) +864 (2512621 - 2045088 \,\zeta_3\big)\Big) \, \epsilon^3 $ \\ 
\hline \vcrule 
$ {g_\lambda}_5$ & $ - \cfrac{1}{3}\cfrac{48\pi^2}{\, d_{c6}}\, \epsilon+\cfrac{4\pi^2(44\, d_c^2+511\, d_c+1122)}{5\, d_{c6}^4}\, \epsilon^2 +
\cfrac{\pi^2}{ 1500 \, d_{c6}^7} \Big(26980\, d_c^5- d_c^4 \big(3968047 - 3468096 \,\zeta_3\big) $ \vspace{-0.1cm} \\ 
& $ \hspace{6.1cm}-26\, d_c^3 \big(1541893 - 1353024 \,\zeta_3\big) + 12\, d_c^2 \big(3653327 -396576 \,\zeta_3\big) $ \\ 
& $ \hspace{6.1cm} + 72\, d_c \big(17311117 - 11943936 \,\zeta_3\big)+864 \big(2987981 - 2181168 \,\zeta_3\big)\Big)\, \epsilon^3$ \\ 
\hline \vcrule 
${\Delta_\mu}_5$ & $\cfrac{24\pi^2}{\, d_{c6}}\, \epsilon-\cfrac{6\pi^2(14\, d_c^2+121\, d_c-138)}{5\, d_{c6}^4}\, \epsilon^2 -
\cfrac{\pi^2}{ 1500 \, d_{c6}^7} \Big(3195\, d_c^5- 20\, d_c^4 \big(259615 - 192456 \,\zeta_3\big) $ \vspace{-0.1cm} \\ 
& $ \hspace{6.7cm}-12\, d_c^3 \big(3948749 - 2848932 \,\zeta_3\big) + 1440\, d_c^2 \big(33334 - 34263 \,\zeta_3\big) $ \\ 
& $ \hspace{6.7cm} + 1296\, d_c \big(907519 - 739692 \,\zeta_3\big) + 23328 \big(78457 - 67716 \,\zeta_3\big)\Big) \,\epsilon^3 $ \\ 
\hline \vcrule 
$ {\Delta_\lambda}_5$ & $ - \cfrac{1}{3}\cfrac{24\pi^2}{\, d_{c6}} \, \epsilon-\cfrac{2\pi^2(6\, d_c^2+119\, d_c+858)}{5\, d_{c6}^4}\, \epsilon^2 -
\cfrac{\pi^2}{1500 \, d_{c6}^7} \Big(10135\, d_c^5+ 10\, d_c^4 \big(259801 - 174312 \,\zeta_3\big) $\vspace{-0.1cm} \\
& $ \hspace{6.9cm}+4\, d_c^3 \big(8149309 - 5609412 \,\zeta_3\big) + 240\, d_c^2 \big(515581 - 345546 \,\zeta_3\big) $ \\ 
& $ \hspace{6.9cm} + 144\, d_c \big(1020203 - 541404 \,\zeta_3\big)+ 3456 \big(53663 - 20169 \,\zeta_3\big)\Big) \,\epsilon^3 $ \\ 
\hline \vcrule 
$ {\eta}_5$ & $ \cfrac{3 }{d_{c6}}\,\epsilon-\cfrac{d_c(2802+d_c(767+60\, d_c)}{40\, d_{c6}^4}\,\epsilon^2 -
\cfrac{d_c}{24000 \, d_{c6}^7} \Big(16000\, d_c^5+ 5 d_c^4 \big(309539 - 152928 \,\zeta_3\big) $\vspace{-0.1cm} \\
& $ \hspace{6.3cm}+4\, d_c^3 \big(1333339 + 261468 \,\zeta_3\big) - 480 \, d_c^2 \big(263197 - 267543 \,\zeta_3\big) $ \\ 
& $ \hspace{6.3cm} - 288\, d_c \big(2968601 - 2664738 \,\zeta_3\big) - 432 \big(3021431 - 2774088 \,\zeta_3\big)\Big) \,\epsilon^3 $ \\ 
\hline \vcrule 
$ {\eta'_5}$ & $ \cfrac{3}{d_{c6}}\,\epsilon-\cfrac{d_c(2442+d_c(707+60\, d_c)}{40\, d_{c6}^4}\,\epsilon^2 -
\cfrac{d_c}{ 12000 \, d_{c6}^7} \Big(8000\, d_c^5+ d_c^4 \big(1377941 - 787968 \,\zeta_3\big) $ \vspace{-0.05cm} \\
& $ \hspace{6.1cm}+ d_c^3 \big(10751771 - 4582008 \,\zeta_3\big) - 6 \, d_c^2 \big(8052667 - 9990216 \,\zeta_3\big) $ \\ 
& $ \hspace{6.1cm} - 36\, d_c \big(16694513 - 14815224 \,\zeta_3\big) - 216 \big(5856221 - 4970808 \,\zeta_3\big)\Big) \, \epsilon^3 $ \\ 
\hline \vcrule 
$ {\phi }_5$ & $ \cfrac{3\,d_c}{2 d_{c6}^3} \epsilon^2 - \cfrac{d_c}{ 8000 d_{c6}^6} \Big(3\, d_c^3 \big(134243 - 90144 \,\zeta_3\big) + 8\, d_c^2 \big(371711 - 222588 \,\zeta_3\big) $ \vspace{-0.3cm} \\ 
& $ \hspace{7.5cm}-252\ d_c \big(31513 - 31104 \,\zeta_3\big) -720 \big(94493 - 73224 \,\zeta_3\big)\Big) \, \epsilon^3 $ \\ 
\hline
\end{tabular}
\end{center}
\vspace{-0.5cm}
\caption{Coordinates of the fixed point $P_5$ and corresponding anomalous dimensions at three-loop order.}
\label{tablexpo1}
\end{table}
%

\begin{table}[h!]
\begin{center}
\begin{tabular}{|c|l|}
\hline \vcrule
$\mu_c$ & $ -\cfrac{8\pi^2}{\, d_{c6}^2}\, \epsilon^2 +
\cfrac{\pi^2}{1500 \, d_{c6}^4} \Big(9\, d_c^2 \big(44081 - 30048 \,\zeta_3\big) + 2\, d_c \big(276557 - 79056 \,\zeta_3\big) -120 \big(92233 - 73224 \,\zeta_3\big)\Big) \,\epsilon^3$ \\ 
\hline \vcrule 
${g_\mu}_c$ & $\cfrac{48\pi^2}{\, d_{c6}}\, \epsilon-\cfrac{4\pi^2(47\, d_c+21)}{5\, d_{c6}^3}\, \epsilon^2 -\cfrac{\pi^2}{3000 \, d_{c6}^5} \Big(3\, d_c^3 \big(195803 - 90144 \,\zeta_3\big) - 8\, d_c^2 \big(942497 - 1026756 \,\zeta_3\big) $ \vspace{-0.25cm} \\ 
& \hspace{6.2cm} $- 2196\, d_c \big(12421 - 12528 \,\zeta_3\big) + 144 \big(1591741 - 1312848\,\zeta_3\big)\Big) \, \epsilon^3 $ \\ 
\hline \vcrule 
$ {g_\lambda}_c$ & $ - \cfrac{1}{3}\cfrac{48\pi^2}{\, d_{c6}} \, \epsilon+\cfrac{4\pi^2(127\, d_c+501)}{15\, d_{c6}^3}\, \epsilon^2 +\cfrac{\pi^2}{ 9000 \, d_{c6}^5} \Big(3\, d_c^3 \big(186203 - 90144 \,\zeta_3\big) - 8\, d_c^2 \big(773567 - 958716 \,\zeta_3\big) $ \vspace{-0.25cm} \\ 
& \hspace{6.5cm} $- 684\, d_c \big(9179 - 30672 \,\zeta_3\big) + 144 \big(2085601 - 1448928\,\zeta_3\big) \Big) \, \epsilon^3 $ \\ 
\hline \vcrule 
${\Delta_\mu}_c$ & $\cfrac{24\pi^2}{\, d_{c6}} \, \epsilon-\cfrac{2\pi^2(32\, d_c-69)}{5\, d_{c6}^3}\, \epsilon^2 -\cfrac{\pi^2}{ 3000 \, d_{c6}^5} \Big(9\, d_c^3 \big(44791 - 30048 \,\zeta_3\big) - 16\, d_c^2 \big(174209 - 167022 \,\zeta_3\big) $ \vspace{-0.25cm} \\ 
& \hspace{6.8cm} $- 1944\, d_c \big(7109 - 5732 \,\zeta_3\big) + 144 \big(706763 - 609444 \,\zeta_3\big) \Big) \, \epsilon^3 $ \\ 
\hline \vcrule 
$ {\Delta_\lambda}_c$ & $ - \cfrac{1}{3}\cfrac{24\pi^2}{\, d_{c6}} \, \epsilon-\cfrac{2\pi^2(28\, d_c+429)}{15\, d_{c6}^3}\, \epsilon^2 +\cfrac{\pi^2}{ 3000 \, d_{c6}^5} \Big(d_c^3 \big(111973 - 90144 \,\zeta_3\big) - 4\, d_c^2 \big(598867 - 452736 \,\zeta_3\big) $ \vspace{-0.25cm} \\ 
& \hspace{6.6cm} $- 8\, d_c \big(2499059 - 1844532 \,\zeta_3\big) - 96 \big(108751 - 40338 \,\zeta_3\big) \Big) \, \epsilon^3 $ \\ 
\hline \vcrule 
$ {\eta}_c= {\eta'_c}$ & $ \cfrac{3}{d_{c6}}\,\epsilon-\cfrac{3 d_c(20\, d_c+149)}{40\, d_{c6}^3}\, \epsilon^2 -\cfrac{d_c}{24000 \, d_{c6}^5} \Big(16000\,d_c^3 + d_c^2 \big(550237 - 223776 \,\zeta_3\big) $ \vspace{-0.25cm} \\ 
& \hspace{6.4cm} $- d_c \big(5283790 - 5670000 \,\zeta_3\big) - \big(47601132 - 42075936 \,\zeta_3\big) \Big) \, \epsilon^3 $ \\ 
\hline
\end{tabular}
\end{center}
\vspace{-0.5cm}
\caption{Coordinates of the fixed point $P_c$ and corresponding anomalous dimensions at three-loop order.}
\label{tablexpo2}
\end{table}

\newpage 

\section{Anomalous dimensions at \texorpdfstring{$P_5$}{p5} and \texorpdfstring{$P_c$}{pc} at order \texorpdfstring{$\epsilon^3$}{ep3} from NPRG approach.}

\vspace{0cm}

We provide here the anomalous dimensions obtained from the NPRG approach of Coquand et al.[22] reexpanded at order $\epsilon^3$. 

\label{NPRG}
\begin{table}[htbp]
\begin{center}
\begin{tabular}{|c|l|}
\hline \vcrule
$\eta_5$ & $ \cfrac{3 }{d_{c6}}\,\epsilon-\cfrac{d_c(4896+1734\, d_c+155\, d_c^2)}{80\, d_{c6}^4}\epsilon^2 + \cfrac{d_c (5375 \, d_c^5- 5178
\, d_c^4-1711125\, d_c^3 - 18385218\, d_c^2-75013452\, d_c-109355832) }{19200\, d_{c6}^7}\epsilon^3$ \\
\hline \vcrule
$\eta'_5$ & $\cfrac{3}{d_{c6}}\,\epsilon-\cfrac{d_c(5544+1962\, d_c+175 \, d_c^2)}{80\, d_{c6}^4}\epsilon^2 + \cfrac{d_c (6475 \, d_c^5- 7914
\, d_c^4-2132397\, d_c^3 - 23019066\, d_c^2- 94606380 \, d_c-138990168) }{19200 \, d_{c6}^7}\epsilon^3$ \\ 
\hline \vcrule
$\eta_c$ & $\cfrac{3}{d_{c6}}\,\epsilon-\cfrac{d_c(2556 +425\, d_c)}{240\, d_{c6}^3}\epsilon^2 +\cfrac{d_c(129925\, d_c^3 -894738\, d_c^2 - 24905043\, d_c-89157186)}{518400\, d_{c6}^5} \epsilon^3$ \\ 
\hline
\end{tabular}
\end{center}
\vspace{-0.5cm}
\caption{Anomalous dimensions $\eta_5$, $\eta'_5$ and $\eta_c$ at order $\epsilon^3$ obtained from the NPRG approach of Coquand {\it et} al. \cite{coquand18}.}
\label{tablexpo3}
\end{table}

\vspace*{-0.5cm}

\twocolumngrid


\end{document}